\documentclass[aps,prl,twocolumn,groupedaddress,
showpacs,amsmath,amssymb]{revtex4}

\usepackage{graphicx}
\usepackage{dcolumn}
\usepackage{bm}

\begin{document}

\title{Electron-Hole Asymmetry in GdBaCo$_{2}$O$_{5+x}$: 
Evidence for Spin Blockade of Electron Transport in 
a Correlated Electron System}

\author{A. A. Taskin}
\author{Yoichi Ando}

\affiliation{Central Research Institute of Electric Power Industry,
Komae, Tokyo 201-8511, Japan}


\begin{abstract}

In $R$BaCo$_{2}$O$_{5+x}$ compounds (R is rare earth) 
variability of the oxygen content allows 
precise doping of CoO$_2$ planes with both types of charge carriers.
We study transport properties of doped GdBaCo$_{2}$O$_{5+x}$ 
single crystals and find a remarkable asymmetry in the behavior of 
holes and electrons doped into a parent insulator 
GdBaCo$_{2}$O$_{5.5}$.
Doping dependences of resistivity, Hall response, and thermoelectric 
power reveal that the doped holes greatly improve the conductivity, 
while the electron-doped samples always remain poorly 
conducting. This doping asymmetry provides strong evidence 
for a spin blockade of the electron transport in $R$BaCo$_{2}$O$_{5+x}$.

\end{abstract}


\maketitle

In condensed matter physics the concept of electron-hole symmetry, 
which states that electrons and holes basically can be regarded as 
equivalent quasiparticles, is one of the fundamental paradigms,
which dramatically simplifies the description of charge transport.
However, in systems with strongly correlated electrons, 
a coupling of spin and charge degrees of freedom can drastically change
this simple picture. One of the most prominent example of how 
spins can affect charge motion is the ``spin blockade'' phenomenon 
observed in quantum dot systems \cite{QDs}.

Recently, existence of a spin blockade for electron transport 
in cobaltites has been suggested in Ref. \cite{Seeb_HBCO} 
to explain the temperature dependence of the thermoelectric power in 
HoBaCo$_{2}$O$_{5.5}$. 
In $R$BaCo$_{2}$O$_{5.5}$ compounds (where $R$ is a rare earth) 
the crystal lattice is composed of equal numbers of CoO$_{6}$ 
octahedra and CoO$_{5}$ square pyramids 
\cite {Seeb_HBCO, Raveau, Akahoshi, Moritomo, Respaud, Frontera, Fauth,
large_GBCO} as schematically shown in Fig. 1(a) and the valence of all the 
Co ions is 3+. 
At low temperature the Co$^{3+}$ ions adopt the low-spin (LS) state in 
octahedral positions and the intermediate-spin (IS) state in pyramidal 
positions \cite {Seeb_HBCO, Raveau, Respaud, Frontera, Fauth, large_GBCO}. 
Generation of electron-hole pairs, or Co$^{2+}$-Co$^{4+}$ states in localized 
picture, determines the transport behavior of $R$BaCo$_{2}$O$_{5.5}$ 
\cite{Seeb_HBCO, Raveau, Akahoshi, Moritomo, Respaud, Frontera, Fauth, 
large_GBCO}. 
Since electron and hole carry not only a charge, but also a spin, their 
motion through a Co$^{3+}$ background can be rather different. In particular, 
it was argued that hopping of a high-spin (HS) Co$^{2+}$ electron through 
a LS Co$^{3+}$ host should be suppressed because of a spin blockade 
mechanism \cite{Seeb_HBCO}, effectively excluding electrons from 
the overall charge transport. A large and positive Seebeck coefficient in 
HoBaCo$_{2}$O$_{5.5}$ observed at low temperatures has been 
suggested in Ref. \cite{Seeb_HBCO} to be a possible evidence of 
the spin blockade for electron transport. 

However, the Seebeck coefficient $S$ in oxygen-variable 
$R$BaCo$_{2}$O$_{5+x}$ is very sensitive to oxygen content; 
it diverges at $x=0.5$, changing sign from negative for $x<0.5$ to positive 
for $x>0.5$ \cite{large_GBCO, proc}. In fact, at the exact $x=0.5$ 
composition the low-temperature Seebeck coefficient 
turns out to be negative \cite{large_GBCO}. Even a small excess of oxygen 
content over $x=0.5$ can provide holes into the system and makes 
$S$ to be positive (it is possible that this was the case with 
HoBaCo$_{2}$O$_{5.5}$ studied in Ref. \cite{Seeb_HBCO}). Therefore, 
to establish the existence of the spin blockade of electron transport, 
one needs to make a direct comparison of electron 
and hole behavior in CoO$_2$ planes .

\begin{figure}[b]
\vspace{-6pt}
\includegraphics*[width=19pc]{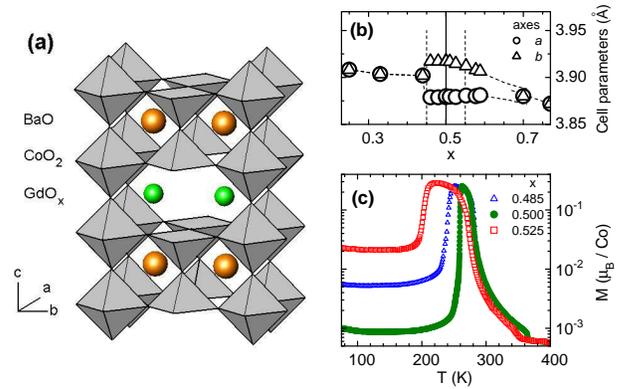}
\caption
{(a) A sketch of the crystal structure of GdBaCo$_{2}$O$_{5.5}$. 
Owing to the ordering of oxygen into alternating 
empty and filled chains, Co$^{3+}$ ions 
become nonequivalent and exhibit either LS state in octahedral 
positions, or IS state in pyramidal positions. 
(b) The evolution of room-temperature in-plane lattice parameters in 
GdBaCo$_{2}$O$_{5+x}$ under variation of oxygen content $x$, 
demonstrating that the orthorhombic crystal structure survives only 
in the narrow range around $x=0.5$.
(c) The magnetization behavior in GdBaCo$_{2}$O$_{5+x}$ 
with oxygen content close to $x=0.5$ 
in a magnetic field of $0.1$ T applied along the $ab$ plane.}
\label{fig1}
\end{figure}

In this Letter, we present a detailed study of doping dependences of 
resistivity, Hall coefficient, and thermoelectric power in 
GdBaCo$_{2}$O$_{5+x}$ single crystals. This study reveals a remarkable 
asymmetry in the behavior of holes and electrons: For hole doping, 
concomitant decrease of the absolute values of resistivity $\rho$, 
Hall coefficient $R_{H}$, and Seebeck coefficient $S$ with increasing 
carrier concentration indicates an eventual establishment of a metallic 
state. On the other hand, for electron doping, the temperature 
dependences of resistivity and Hall coefficient do not show any appreciable 
change, indicating that electrons introduced into the system are effectively 
immobile. This doping asymmetry strongly supports the idea that 
the spin blockade of electron transport takes place in 
$R$BaCo$_{2}$O$_{5+x}$, making a solid case that the spin blockade 
phenomena occurs not only in quantum dot systems but also 
in transition-metal oxides.

High-quality GdBaCo$_{2}$O$_{5+x}$ single crystals are grown using 
a floating-zone technique \cite{large_GBCO}. Carrier doping to single 
crystal samples is performed by precisely changing the oxygen content 
$\Delta x$ with an accuracy better than 0.001 using an elaborate set of 
high-temperature annealing and quenching procedures 
\cite{large_GBCO, ox_diff}. To avoid complications due to a modification 
of the crystal structure upon changing the oxygen concentration 
\cite{Akahoshi, large_GBCO}, the change of oxygen content in the present 
study is restricted to the range of $0.45<x<0.55$, where 
GdBaCo$_{2}$O$_{5+x}$ keeps an orhorhombic crystal structure 
originating from anion ordering 
as evidenced by the doping dependence of the in-plane lattice 
parameters shown in Fig 1(b). In addition, in this narrow doping range the dc 
magnetization in GdBaCo$_{2}$O$_{5+x}$ [see Fig. 1(c)] 
shows qualitatively the same type of successive paramagnetic (PM) to 
ferromagnetic (FM), and then to antiferromagnetic (AF) phase transitions upon 
cooling in both parent and doped crystals, implying basically 
the same underlying microscopic spin structure \cite{spin_str}. 
Since oxygen vacancies created in the GdO$_{x}$ plane upon doping
do not change the in-plane Co-O framework [see Fig. 1(a)], all transport
measurements have been performed along the $\it{ab}$ plane.

The resistivity is measured using a standard ac four-probe method. 
Both current and voltage contacts are made with gold paint before all the heat 
treatment procedures that are used to vary the oxygen content. The Hall 
resistivity is measured using a standard six-probe technique by sweeping 
the magnetic field ${\bf H}$$\,\parallel\,$$c$ to both plus and minus 
polarities at fixed temperatures; the electric current is always along 
the $ab$-plane. The thermoelectric power is measured in a slowly oscillating 
thermal gradient of $\sim 1$ K along the $ab$ plane. The contribution from the 
gold wires ($\sim 2$ $\mu$V/K) used as output leads is subtracted. 

\begin{figure}[t]
\includegraphics*[width=19pc]{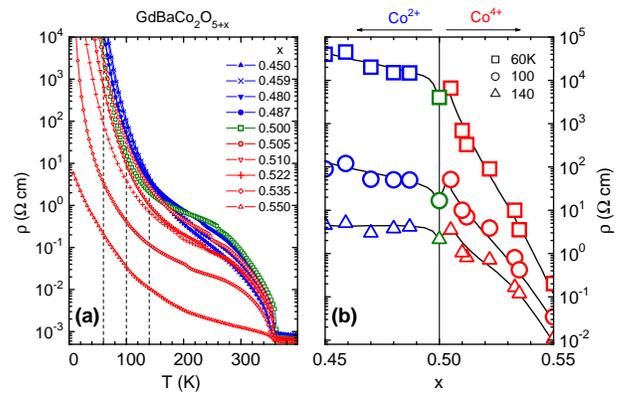}
\caption{(a) Temperature dependences of the in-plane resistivity 
$\rho(T)$ of GdBaCo$_{2}$O$_{5+x}$
crystals with oxygen content close to $x=0.5$. 
(b) Doping dependences of the in-plane resistivity $\rho(x)$ 
of GdBaCo$_{2}$O$_{5+x}$ crystals at several temperatures 
(shown by dashed lines in the left panel).}
\label{fig2}
\end{figure}

Figure 2(a) shows temperature dependences of the in-plane resistivity $\rho$ 
of GdBaCo$_{2}$O$_{5+x}$ crystals for $0.45<x<0.55$. At high temperatures 
all curves show an almost temperature-independent ``metallic" 
behavior. Upon cooling below $\approx$ 360 K GdBaCo$_{2}$O$_{5+x}$ 
crystals undergo a metal-insulator transition (MIT).
As has been established for $R$BaCo$_{2}$O$_{5.5}$ 
\cite{Raveau, Akahoshi, Moritomo, Respaud, Frontera},
the MIT coincides with a spin-state transition of Co$^{3+}$ ions: 
On the metallic side, a half of cobalt ions (in octahedral positions) adopts 
the HS state and another half (in pyramidal positions) adopts the IS state. 
On the insulator side HS-Co$^{3+}$ ions change their spin state into the LS 
state \cite{Seeb_HBCO, Raveau, Frontera, Fauth, large_GBCO, spin_str}. 
This spin-state transition is manifested in the change of 
the magnetization behavior \cite{Respaud, large_GBCO, spin_str} 
at $\approx$ 360 K [see Fig 1(c)].

Since we are interested in how doping changes the ground state of a parent 
insulator, the focus of our study is on the doping dependence of transport 
properties at low temperatures. For both electron-doped and hole-doped crystals 
the $\rho(T)$ curves show clear kinks at $T_N$, where the FM ordering abruptly 
changes into an AF one [see Fig 1(c)]. 
In the AF state the resistivity of GdBaCo$_{2}$O$_{5+x}$ 
grows rapidly with decreasing temperature. Although all $\rho(T)$ curves 
demonstrate essentially insulating behavior in the $0.45<x<0.55$ concentration 
range, an introduction of carriers into the system changes the absolute 
value of resistivity, and this change is very different for electrons and holes.
The difference in the doping response is most clearly illustrated in Fig. 2(b), 
which shows the doping dependence of resistivity in the AF phase at several 
temperatures [shown by dashed lines in the Fig 2(a)]. The most striking feature 
here is an asymmetry for electron and hole doping: 
While the hole doping leads to a steady decrease in resistivity (for example, 
$\rho$ changes by almost five orders of magnitude at 60 K 
as $x$ increases from 0.50 to 0.55), 
the electron doping causes little change in resistivity. 

\begin{figure}[t]
\includegraphics*[width=19pc]{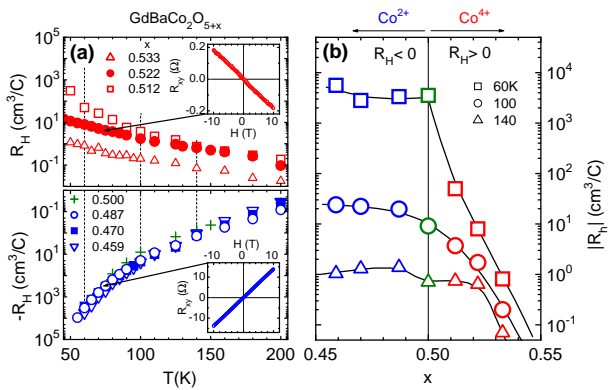}
\caption{(a) Temperature dependences of the Hall coefficient 
$R_{H}(T)$ of GdBaCo$_{2}$O$_{5+x}$
crystals with oxygen content close to $x=0.5$. 
Asymmetric in magnetic field parts of the $R_{xy}(H)$, 
which directly determine a sign and a value of the 
Hall coefficient, are presented in the upper 
and the lower insets. 
(b) Doping dependences of the absolute value 
of the Hall coefficient $|R_{H}(x)|$ 
of GdBaCo$_{2}$O$_{5+x}$ crystals at several temperatures 
(shown by dashed lines in the left panel).}
\label{fig3}
\end{figure}

The same asymmetry in the doping dependence is observed 
in the Hall data as well.
Figure 3(a) shows temperature dependences of the Hall coefficient $R_{H}(T)$ 
in GdBaCo$_{2}$O$_{5+x}$ with oxygen concentrations in the same range as 
for the resistivity data. All measurements are done below $T_N$ in the AF 
phase, where the anomalous Hall effect is expected to be negligible and 
the Hall coefficient has a simple physical meaning. Indeed, as shown in 
the insets of Fig. 3(a), the Hall resistivity is almost perfectly linear in magnetic 
field. Also, as expected for positive 
and negative charge carriers, the sign of the Hall coefficient in 
GdBaCo$_{2}$O$_{5+x}$ is positive for hole-doped crystals ($x>0.5$) and 
negative for electron-doped crystals ($x<0.5$). 
Figure 3(b) shows the doping dependence 
of the absolute value of the Hall coefficient $|R_{H}(x)|$ at the same 
temperatures as for the resistivity data. An asymmetry in the Hall response 
for hole-doped and electron-doped crystals correlates well with the asymmetry 
in the doping dependence of the resistivity: While the Hall coefficient 
decreases by several orders of magnitude upon doping by holes, 
meeting the expectation for a doping procedure, the electron 
doping causes almost no change in the Hall response at any temperature below 
$T_N$ as if electrons introduced into a crystal are immobile.

The thermoelectric power is another transport property that can distinguish 
between electrons and holes. Figure 4(a) shows temperature dependences of the 
Seebeck coefficients in GdBaCo$_{2}$O$_{5+x}$. At high temperatures $S$ is very 
small and negative for both electron-doped and hole-doped crystals, suggesting 
that the doping little affects the ``metallic" state. On the other hand, at low 
temperatures the thermoelectric power is very sensitive to the type of carriers: 
$S>0$ for hole-doped crystals, and $S<0$ for electron-doped counterparts, 
similar to the Hall coefficient behavior. Note that the temperature 
dependences of the Seebeck coefficient $S(T)$ do not follow a simple law 
related to an insulating behavior neither for hole-doped nor 
for electron-doped crystals. 

\begin{figure}[t]
\includegraphics*[width=19pc]{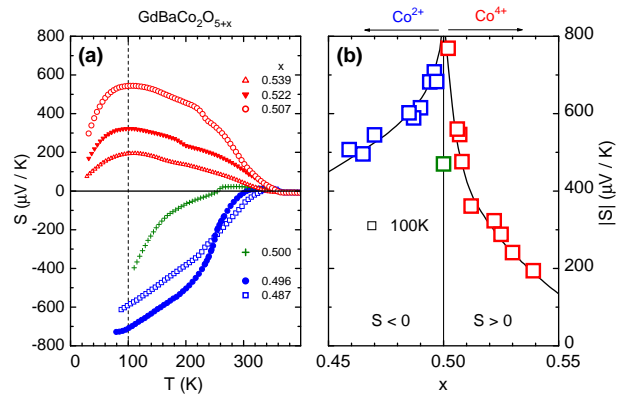}
\caption{(a) Temperature dependences of the Seebeck coefficient 
$S(T)$ of GdBaCo$_{2}$O$_{5+x}$
crystals with oxygen content close to $x=0.5$. 
(b) The doping dependence of the absolute value 
of the Seebeck coefficient $|S(x)|$ 
in GdBaCo$_{2}$O$_{5+x}$ at $T=100$ K.}
\label{fig4}
\end{figure}

The most prominent feature of the doping dependence of the Seebeck coefficient, 
shown in Fig. 4(b) for $T=100$ K, is a sharp divergence at $x=0.5$. Its absolute 
value $|S|$ is extremely large in the immediate proximity to $x=0.5$, reaching 
$\sim$800 $\mu$V/K, and decreases rapidly as $x$ deviates from 0.5, albeit 
in a different way for electrons and holes. Such $S(x)$ behavior has been 
attributed to the hopping transport of charge carriers in GdBaCo$_{2}$O$_{5+x}$ 
\cite{proc}. 

Thus, as evidenced by the resistivity, Hall effect, and thermoelectric power 
measurements, both electrons and holes can be introduced into the parent 
GdBaCo$_{2}$O$_{5.5}$ by changing its oxygen content, but their transport 
is very different. 

In order to understand the origin of the electron-hole asymmetry in this 
material, one needs to understand the mechanism of charge transport 
in GdBaCo$_{2}$O$_{5+x}$ at low temperature. 
The parent GdBaCo$_{2}$O$_{5.5}$ is a gapped insulator as 
evidenced by an activation behavior of conductivity [Fig. 5(a)]. Excitations of 
electron-hole pairs across the energy gap $\approx 140$ meV determine the 
temperature dependences of conductivity, Hall coefficient, and thermoelectric 
power \cite{large_GBCO}. The low-temperature transport in the parent material 
is dominated by electrons, which are thermally created in relatively wider 
energy bands than holes; 
the higher mobility of electrons determines the negative sign of the Hall 
coefficient and the Seebeck coefficient \cite{note}.  The concentration of 
electrons (and holes) in GdBaCo$_{2}$O$_{5.5}$ decreases exponentially 
with decreasing temperature [down to $\sim 10^{15}$ cm$^{-3}$ 
at 60 K as estimated from the Hall data (Fig. 3)]. 
The ``effective'' mobility $\mu_{H}=\sigma R_{H}$, which would be the Hall 
mobility if only one type of charge carriers were present, 
is around $\sim 1$ cm$^2$V$^{-1}$s$^{-1}$, being already somewhat close 
to a borderline for a ``band'' transport.

A change of oxygen content by only $\Delta x=\pm0.001$ provides a large 
number of extra carriers $\sim 10^{19}$ cm$^{-3}$, implying a great increase 
in conductivity. On the other hand, 
doping of both electrons and holes leads to a diminishing mobility 
[see inset in Fig. 5(a)], suggesting an effective carrier 
localization, which is probably caused by disorders in the oxygen sublattice. 
The conductivity data, shown in Fig. 5(b), reveal a 3D variable-range hopping 
(VHR) behavior of charge carriers in GdBaCo$_{2}$O$_{5+x}$ crystals for both 
electrons and holes. This type of behavior is well known in systems where 
the low-temperature transport is governed by phonon-assisted tunneling of 
carriers between localized states randomly distributed in energy and 
position \cite{Mott}. The increase in the density of states upon doping in 
such systems naturally leads to an increasing conductivity, as is observed 
in hole-doped GdBaCo$_{2}$O$_{5+x}$ crystals for $x>0.5$.
What is peculiar here is that electron doping little changes the conductivity 
of GdBaCo$_{2}$O$_{5+x}$ for $x<0.5$.

\begin{figure}[t]
\includegraphics*[width=19pc]{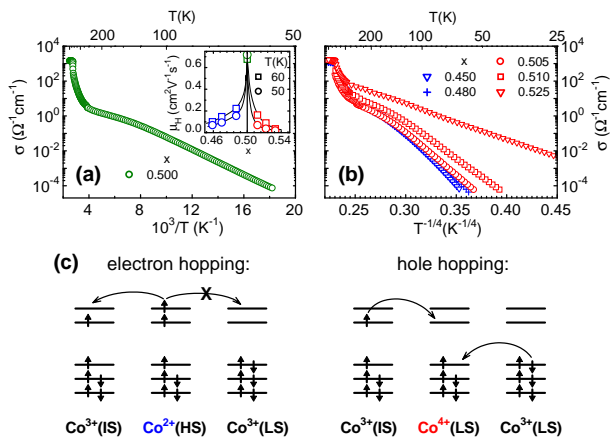}
\caption{(a) The low-temperature conductivity $\sigma(T)$ in the parent 
GdBaCo$_{2}$O$_{5.5}$ compound ($x=0.5$) shows an activation behavior. 
The inset demonstrates a diminishing Hall mobility 
$\mu_{H}=\sigma R_{H}$ caused by a doping-induced disorder.
(b) The low-temperature conductivity $\sigma(T)$ in GdBaCo$_{2}$O$_{5+x}$ 
shows 3D variable-range hopping for crystals doped both by electrons and holes. 
(c) The scheme of electron Co$^{2+}$ and hole Co$^{4+}$ hopping 
through the IS and LS states of Co$^{3+}$ ions, 
illustrating the phenomenon of the spin blockade for electron hopping.}
\label{fig5}
\end{figure}

Keeping in mind the hopping character of the low-temperature charge transport 
in GdBaCo$_{2}$O$_{5+x}$, the difference in the behavior of electrons 
(Co$^{2+}$ ions, moving in the background of Co$^{3+}$ ions) and holes 
(Co$^{4+}$ ions) can be understood as a difference in their hopping probability. 
The motion of an extra carrier through a crystal can be considered as 
a hopping process, which leads to an interchange of the spatial 
positions of a Co$^{2+}$ (or Co$^{4+}$) ion with a neighboring Co$^{3+}$ ion. 
Indeed, if one of the seven $3d$-electrons of a Co$^{2+}$ ion moves to 
a neighboring Co$^{3+}$ ion, which has six $3d$-electrons, 
as shown in the left panel of the Fig. 5(c), the ions interchange their valence 
states, i.e. the Co$^{2+}$ ion moves to the neighboring position. The same is 
true for hopping of a Co$^{4+}$ ion, with the only difference that an electron 
moves from a Co$^{3+}$ ion to a Co$^{4+}$ ion, which has five $3d$-electrons, 
as shown in the right panel of the Fig. 5(c). 
As can be inferred from our discussion for the parent material, 
in GdBaCo$_{2}$O$_{5+x}$ at low temperatures 
nearly half of the Co$^{3+}$ ions are in a LS state and the rest are in 
a IS state. On the other hand, Co$^{2+}$ ions are always in a HS state owing 
to a weaker crystal field than for Co$^{3+}$ ions, and Co$^{4+}$ ions are 
always in a LS state owing to a stronger crystal field \cite{Huheey}.
Figure 5(c) shows the scheme of hopping of HS-Co$^{2+}$ and LS-Co$^{4+}$ 
ions onto IS-Co$^{3+}$ and LS-Co$^{3+}$ ions: As one can see, a LS-Co$^{4+}$ 
ion can always interchange its position with an IS-Co$^{3+}$ ion or 
a LS-Co$^{3+}$ ion, keeping the same spin states before and after the hopping 
event [as shown in the right panel of the Fig. 5(c)]. On the contrary, 
a HS-Co$^{2+}$ ion can interchange its position only with a IS-Co$^{3+}$ ion 
because moving one $3d$-electron to a LS-Co$^{3+}$ ion creates 
a non-HS-Co$^{2+}$ ion which is energetically unfavorable.
This means that hopping through a LS-Co$^{3+}$ ion [marked by 
a cross in the left panel in the Fig. 5(c)] hardly occurs. 
This phenomenon, which has been called the spin blockade \cite{Seeb_HBCO}, 
can effectively supress electron transport in GdBaCo$_{2}$O$_{5+x}$ compounds. 
Hence the concept of the spin blockade provides a simple explanation 
for the observed electron-hole asymmetry in GdBaCo$_{2}$O$_{5+x}$, 
although the rigorous quantitative understanding of the doping dependences of 
the transport properties in GdBaCo$_{2}$O$_{5+x}$ is a challenging issue. 

\begin{acknowledgments}
We thank A. N. Lavrov for fruitful discussions.
\end{acknowledgments}


\begin{thebibliography}{99}

\bibitem{QDs} For a example, see D. Weinmann, W. H\"{a}usler, and 
B. Kramer, Phys. Rev. Lett. {\bf 74}, 984 (1995); 
T. Fujisawa, D. G. Austing, Y. Tokura, Y. Hirayama, and S. Tarucha, 
{\it ibid.} {\bf 88}, 236802 (2002);
K. Ono, D. G. Austing, Y. Tokura, and S. Tarucha, 
Science {\bf 287}, 1313 (2002);

\bibitem{Seeb_HBCO} A. Maignan, V. Caignaert, B. Raveau, D. Khomskii,
and G. Sawatzky, Phys. Rev. Lett. {\bf 93}, 026401 (2004).

\bibitem{Raveau} C. Martin, A. Maignan, D. Pelloquin, N. Nguyen, and B. Raveau, 
Appl. Phys. Lett. {\bf 71}, 1421 (1997); A. Maignan, C. Martin, D. Pelloquin, 
N. Nguyen, and B. Raveau, 
J. Solid State Chem. {\bf142}, 247 (1999).

\bibitem{Akahoshi} D. Akahoshi and Y. Ueda,
J. Solid State Chem. {\bf 156}, 355 (2001).

\bibitem{Moritomo} Y. Moritomo {\it et al.},
Phys. Rev. B {\bf 61}, R13325 (2000).

\bibitem{Respaud} M. Respaud {\it et al.},
Phys. Rev. B {\bf 64}, 214401 (2001).

\bibitem{Frontera} C. Frontera, J. L. Garc\'{i}a-Mu\~{n}oz, A. Llobet, 
and M. A. G. Aranda, 
Phys. Rev. B {\bf 65}, 180405(R) (2002).

\bibitem{Fauth} F. Fauth, E. Suard, V. Caignaert, and I. Mirebeau,
Phys. Rev. B {\bf 66}, 184421 (2002).

\bibitem{large_GBCO} A. A. Taskin, A. N. Lavrov, and Y. Ando,
Phys. Rev. B {\bf 71}, 134414 (2005).

\bibitem{proc} A. A. Taskin, A. N. Lavrov, and Y. Ando,
Proc. ICT'03, IEEE, La Grande-Motte, France, 196 (2003).

\bibitem{ox_diff} A. A. Taskin, A. N. Lavrov, and Y. Ando,
Appl. Phys. Lett. {\bf 86}, 091910 (2005).

\bibitem{spin_str} A. A. Taskin, A. N. Lavrov, and Y. Ando,
Phys. Rev. Lett. {\bf 90}, 227201 (2003).

\bibitem{note} The positive Seebeck coefficient of HoBaCo$_{2}$O$_{5.5}$, 
considered in Ref. \cite{Seeb_HBCO} as an evidence of an electron transport 
blockade, might have originated from a slight excess of the oxygen content 
over 5.5. 

\bibitem{Mott} N. F. Mott and E. A. Davis, {\it Electronic Processes in
Non-Crystalline Materials}, 2nd ed. (Clarendon Press, Oxford, 1979).

\bibitem{Huheey} J. E. Huheey, E. A. Keiter, and R. L. Keiter,
{\it Inorganic Chemistry: Principles of structure and reactivity, 4th
edn.} (Harper Collins, New York, 1993).



\end{thebibliography}
\end{document}